\documentstyle[osa,manuscript,psfig]{revtex} 

\begin{document}
\titlepage

\title{
Formation of color-singlet gluon-clusters and inelastic
diffractive scattering\\
{\small \bf Part II. Derivation of the $t$- and $M_x^2/s$-dependence of  
cross-sections in the SOC-approach}}

\author{T. Meng, R. Rittel, K. Tabelow, and Y. Zhang
\thanks{Present address: Chinese Academy of Sciences, Institute 
of Theoretical Physics, POB 2735, Beijing 100080, China.}\\
{\it Institut f\"ur Theoretische Physik, 
Freie Universit\"at Berlin, 
14195 Berlin, Germany} \protect\\
{\small meng@physik.fu-berlin.de}}
\maketitle

\begin{abstract}
The $t$-dependence and the $(M_x^2/s)$-dependence of the double differential
cross-sections for inelastic diffractive 
scattering off proton-target are discussed. Here $t$ stands for the
four-momentum-transfer squared, $M_x$ for the missing mass, and $\sqrt{s}$ for
the total c.m.s. energy. 
It is shown, that the space-time properties of the color-singlet
gluon-clusters due to SOC, discussed in Part I, lead to simple analytical 
formulae for $d^2\sigma/dt\,d(M_x^2/s)$  and  for $d\sigma/dt$, and
that the obtained results
are in good agreement with 
the existing data. Further experiments are suggested.
\end{abstract}


\newpage

\section{Introduction together with a brief summary of Part I}
\label{intro}
It is suggested in the preceding paper\cite{SOCI} 
(hereafter referred to as Part I)
that concepts and methods of Statistical Physics 
for open dynamical complex systems far from equilibrium
can be used in describing the
phenomena  associated with large rapidity gap (LRG) events in
deep-inelastic electron-proton 
scattering\cite{LRGDiscovery,LRGDISatHera} 
and other inelastic 
diffractive scattering 
processes\cite{gammapdata,ppdata,ppbardata,hhdata}. 
This approach is motivated by the following observations:

First, LRG events in
deep-inelastic electron-proton scattering
have been observed\cite{LRGDiscovery,LRGDISatHera}  in the small-$x_B$ region 
($x_B$$<$$ 10^{-2}$, say) --- a kinematic region in which
low-energy (soft) gluons dominate\cite{GluonDominance}. 
This piece of experimental facts indicates that
the occurrence of LRG events and the observed
gluon-domination in this region are very much related to each other.
 
Second, the characteristic properties of the gluons --- in particular the
local gluon-gluon coupling prescribed by the QCD-Lagrangian, the
confinement, and the non-conservation of gluon-numbers --- strongly
suggest that systems of interacting soft gluons should be considered
as {\em open, dynamical, complex} systems with {\em many degrees of
freedom,}  and
that such  systems are  in general {\em far from equilibrium}.

Third, it has been proposed by Bak, Tang and Wiesenfeld 
(BTW)\cite{OriginalSOC} 
some time ago, that a wide class of 
open dynamical complex systems far from equilibrium  may evolve in a 
self-organized manner to critical states, where perturbations caused 
by local interactions can initiate long-range-correlations through
 ``domino effects'' 
which lead to spatial and temporal 
fluctuations extending
over all length and time scales --- in form of avalanches
(which we call BTW-avalanches in Part I and hereafter).
The size-distributions and the lifetime-distributions of such
ava\-lanches exhibit {\em power-law behaviors}.
These behaviors are {\em universal} and {\em robust}.
In fact these behaviors are 
considered as {\em the fingerprints of self-organized criticality}
(SOC). 
In the macroscopic world, there 
are many open dynamical complex systems which show this kind of 
power-law  
behaviors\cite{OriginalSOC,ReviewSOC}: 
sandpiles, earthquakes, wood-fire, evolution, traffic jam, stock
market etc. 

Having these observations in mind,
we are automatically led to the questions:
Can self-organized criticality also exist  
in the microscopic world --- at the level of gluons and quarks?
In particular, can there be BTW-avalanches in systems of interacting
soft gluons?
Is it possible to probe the existence and the properties of such avalanches
experimentally --- by examing reactions in which the interactions of
soft gluons play a significant role?

To answer these questions, it is useful to recall the following: Inelastic
diffractive scattering processes are characterized (see e.g. Gallo in 
Ref.[\ref{SmalltNew}] and the papers cited therein) by 
large rapidity gaps in the final state and the
existence of such gaps has been interpreted as the consequence of
``the exchange of colorless objects.'' These objects can be, and have been
\cite{softgluon}, associated with color-singlet systems/clusters of 
interacting soft gluons. 
Such colorless objects can exist inside and/or outside
the proton, and the interactions between such color-singlets, as well as 
those between such objects and ``the mother proton'', should be of Van
der Waals type. Hence it is expected that such colorless objects can be
easily separated from the other color-singlets including
``the mother proton'' in peripheral collision processes, in which (in
contrast to ``hard scattering'') not much transfer of momentum is
needed. Furthermore, since the process of ``carring away'' or
``knocking-out'' of such a colorless object from the proton is
comparable with the knocking-out of a nucleon off a nucleus by
energetic beam-particles, it is also expected that the characteristic
features (which does not include the absolute values of the cross-sections) of 
such inelastic diffractive processes should be {\em independent} of
the incident energy, and {\em independent} of the quantum-numbers of
the projectile. 

Based on these knowledge and expectations, 
we performed a systematic analysis\cite{SOCI} of
the existing data for inelastic diffractive scattering in
electron-proton scattering processes \cite{LRGDiscovery,LRGDISatHera}, 
in photoproduction \cite{gammapdata}, 
and in proton-proton and proton-antiproton 
collisions \cite{ppdata,ppbardata,hhdata}. 
The obtained results (which
are presented in Part I) can be summarized as follows:

The data 
\cite{LRGDiscovery,LRGDISatHera,gammapdata,ppdata,ppbardata,hhdata}
show
that the above-mentioned characteristic 
features for SOC {\em indeed exist},
and that the relevant exponents 
in such power-laws
are approximately the {\em same}
for {\em different} reactions.
The observed features imply that a color-singlet gluon-cluster is a
BTW-avalanche, and as such, it can have 
{\em neither a typical size, nor a typical lifetime, nor
a given static structure}. In fact,
it  has much in common with an earthquake, or an avalanche of 
snow (see Sections 2, 4 and 5 of Part I).
By examing the data for inelastic diffractive scattering 
processes performed at different incident energies and/or 
in which different kinds of beam-particles are used, 
we are able to extract information about the colorless objects without 
specifying their structures. Furthermore, since the obtained results are
approximatly independent of the incident energy and independent of 
the quantum numbers of the projectile, we can conclude
that the extracted knowledge about the
color-singlet gluon-clusters are universal.
Hence, the following picture emerges:

Viewed from the beam-particle, the target-proton in
diffractive scattering appears as a ``cloud'' of colorless 
objects which in general exist partly inside and partly outside the 
confinement region of the proton. Such objects are
color-singlet gluon-clusters
which exhibit the characteristic features of
BTW-avalanches as a consequence of SOC. 
In (geometrically speaking) more peripheral scattering processes, such as
inelastic diffractive scattering, the
beam-particle will encounter one of these color-singlet gluon-clusters 
and ``carry it away'', because the interaction between the
struck object and any other neighboring color-singlets (including the
``mother proton'') are expected to be of Van der Waals' type.

We note, in Part I of this paper,
we simply adopted the currently popular definition of
``inelastic diffractive scattering processes''. That is, 
when we talked about ``inelastic diffractive scattering'' we
were always referring to processes in which
``colorless objects'' are ``exchanged''. 
In other words, in that part of the paper,
the following question has {\em not} been asked: 
Are the above-mentioned ``inelastic {\em diffractive} scattering
processes'' indeed comparable with {\em diffraction} in
Optics, in the sense that the beam particles should be considered
as waves, and the target-proton together with the associated 
(in whatever manner) colorless objects can indeed be viewed as a 
``scattering screen''? 

In Part II of this paper, we discuss in detail this question, and we
examine in particular the existing data \cite{ppdata,ppbardata} for 
the double differential cross-section $d^2\sigma/dt\,d(M_x^2/s)$ for 
proton-proton and antiproton-proton collisions 
(where $t$ is the 
4-momentum-transfer squared, $M_x$ is the missing-mass, and $\sqrt{s}$ is the 
total c.m.s. energy). To be more precise,
what we wish to find out in this connection is: ``Can the observed 
$t$-dependence and the $(M_x^2/s)$-dependence of  $d^2\sigma/dt\,d(M_x^2/s)$ 
in the given kinematic range  
($0.2\mbox{\,GeV}^2$$\le$$|t|$$\le$ $3.25 \mbox{\,GeV}^2$,
$16\mbox{\,GeV}$$\le$$\sqrt{s}$$\le$ $630\mbox{\,GeV}$, 
and $M_x^2/s$$\le$ $0.1$)
be understood in terms of the well-known concept of diffraction in Optics? 

The answer to this question is of particular interest for several
reasons:

(a) High-energy proton-proton and proton-antiproton scattering at
small momentum transfer has played, and is still playing a very
special role in understanding diffraction
and/or diffractive dissociation in lepton-, photon-
and hadron-induced 
reactions\cite{LRGDiscovery,LRGDISatHera,gammapdata,ppdata,ppbardata,hhdata,ReviewSmallt,SmalltNew,OpticalModels,ReggeModels}. 
Many
experiments
have been
performed at various incident energies for elastic and inelastic
diffractive scattering processes. It is known that the double
differential cross section $d^2\sigma/dt\,d(M_x^2/s)$ is a quantity
which can yield much information on the reaction
mechanism(s) and/or on the structure of the participating colliding 
objects. In the past, the 
$t$-, $M_x$- and $s$-dependence of the differential cross-sections for
inelastic diffractive scattering processes
has been presented in different forms,
where a number of interesting features have been 
observed\cite{ppdata,ppbardata,ReviewSmallt}.
For example, it is seen that, the $t$-dependence of 
$d^2\sigma/dt\,dM_x^2$ at fixed $s$
depends very much on $M_x$; 
the $M_x^2$-dependence of 
$d^2\sigma/dt\,dM_x^2$
at fixed $t$ depends on $s$.
But, when $d^2\sigma/dt\,d(M_x^2/s)$ is plotted as
function of $M_x^2/s$ at 
given $t$ -values (in the range 
$0.2\,\mbox{GeV}^2\le |t|\le 3.25\,\mbox{GeV}^2$)
they are approximately independent of
$s$! What do these observed striking features tell us?
The first precision
measurement of this quantity was published more than twenty years 
ago\cite{ppdata}. 
Can this, as well as the more recent
$d^2\sigma/dt\,d(M_x^2/s)$-data\cite{ppbardata} be understood theoretically? 

(b) The idea of using optical and/or geometrical analogies to describe 
high-energy hadron-nucleus and hadron-hadron collisions at small
scattering angles has been discussed by many authors
\cite{OpticalModels,ReviewSmallt} many years ago. It is shown in
particular that this approach is very useful and 
successful in describing elastic scattering.
However, it seems that, until now, no attempt has been made 
to describe the  data\cite{ppdata,ppbardata} by
performing 
quantitative calculations for 
$d^2\sigma/dt\,d(M_x^2/s)$
by using optical geometrical analogies.
It seems worthwhile to make such an attempt. This is because, it
has been pointed out\cite{SOCLetter} very recently, that 
the above-mentioned analogy
can be made to understand the observed $t$-dependence in
$d\sigma /dt$.

(c) Since inelastic diffractive $pp$- and $\bar{p}p$-scattering
belongs to those soft processes in which the initial and final states
are well-known hadrons, it is expected that also they 
can be described in the well-known Regge-pole approach
\cite{ReggeModels,ReviewSmallt,SmalltNew}. 
The basic idea of this approach is that colorless objects in form of
Regge trajectories (Pomerons, 
Reggions etc.) are exchanged during the collision,
and such trajectories are responsible for the dynamics of the
scattering processes.
In this approach, it is
the $t$-dependence of the
Regge trajectories, the $t$-dependence 
of the corresponding
Regge residue functions, the properties of the coupling of the
contributing trajectories (e.g. triple Pomeron or  
Pomeron-Reggion-Pomeron coupling),
and the number of contributing Regge trajectories which
determine the experimentally observed $t$- and  $M_x$-dependence of
$d^2\sigma/dt\,d(M_x^2/s)$.  
A number of Regge-pole models\cite{SmalltNew,ReggeModels} 
have been 
proposed, and there exist good fits \cite{SmalltNew,ReggeModels} to the 
data. What remains to be understood in this approach is the dynamical
origin of the Regge-trajectories on the one hand, and the physical
meaning of the unknown functions (for example the $t$-dependence of
any one of the Regge-residue functions) on the other.
It has been pointed out \cite{Sterman,SOCLetter}, that there may 
be an overlap between the ``Partons in Pomeron and Reggeons'' picture
and the SOC-picture \cite{SOCLetter}, and that one way 
to study the possible relationship between the two approaches 
is to take a closer look at the double differential cross-section
$d^2\sigma/dt\,d(M_x^2/s)$.

\section{Optical diffraction off dynamical complex systems}
\label{sec:2}
Let us begin our discussion on the above-mentioned questions by
recalling that the concept of ``diffraction'' or ``diffractive scattering''
has its origin in Optics, and Optics is part of Electrodynamics, 
which is not only the {\em classical limit}, but also {\em the
basis} of Quantum Electrodynamics (QED).
Here, it is useful to recall in particular the following:
Optical diffraction is associated with 
departure from geometrical optics 
caused by the finite wavelength of light.
Frauenhofer diffraction can be observed by placing  
a scatterer (which can in general be a
scattering screen with more than one aperture or a 
system of scattering objects) 
in  the path of propagation of light (the wavelength 
of which is less than the linear dimension of the scatterer) where 
not only
the light-source, but also the detecting device, are very far 
away from the 
scatterer. The parallel incident light-rays can be 
considered as plane waves (characterized by a set of constants 
$\vec k, w\equiv |\vec k|$, and $u$ say, which denote
the wave vector, the frequency and the  amplitude of a component of the  
electromagnetic field respectively in the laboratory frame).  
After the scattering, 
the scattered field can be written in accordance with Huygens' principle as
\begin {equation} \label{1}
u_P = \frac{e^{i|\vec{k}^\prime|R}}{R} f(\vec{k},\vec{k}^\prime).
\end{equation}
Here, $u_P$ stands for a component of the field originating from 
the scatterer, $\vec{k}^\prime$ is the wave vector of the scattered light 
in the direction of observation, $|\vec{k}^\prime|\equiv \omega^\prime$ 
is the corresponding frequency, $R$ is the distance between the 
scatterer and the observation point $P$, and 
$f(\vec{k},\vec{k}^\prime$) is the (unnormalized) scattering 
amplitude which describes the change of the wave vector in the 
scattering process. By choosing a coordinate system in which the 
$z$-axis coincides with the incident wave vector $\vec{k}$, 
the scattering amplitude can be expressed as 
follows\cite{Landau,OpticalModels,ReviewSmallt}
\begin{eqnarray}\label{2} f(\vec q) & = & 
\frac{1}{(2\pi)^2} \int\!\int\limits_{\mbox{\hspace*{-0.5cm}}\Sigma}^{} 
              d^2\vec{b}\,\alpha(\vec{b})\,
              e^{-i\vec q \cdot \vec{b}}\mbox{\ .}
\end{eqnarray}
Here,  $\vec{q}\equiv \vec{k}^\prime - \vec{k}$ determines 
the change in wave vector due to diffraction; $\vec{b}$ is the 
impact parameter which indicates the position of 
an infinitesimal surface element on the wave-front ``immediately
behind the scatterer'' where the incident wave would reach in the
limit of geometrical optics, and $\alpha(\vec{b})$ 
is the corresponding amplitude 
(associated with the boundary conditions which the scattered field
should satisfy)
in the two-dimensional
impact-parameter-space (which is here the $xy$-plane), and the
integration extends over the region $\Sigma$ in which
$\alpha(\vec{b})$ is different from zero. In those cases in which the 
scatterer is symmetric with respect to the scattering axis (here the $z$-axis),
Eq.(2) 
can be expressed, by using an integral representation 
for $J_0$, as
\begin{eqnarray}
f(q) & = & \frac{1}{2\pi} \int\limits_{0}^{\infty} b\, db\, 
                 \alpha(b) J_0(q b)\mbox{\ .}
\end{eqnarray}
where $q$ and $b$ are the magnitudes of $\vec{q}$ and
$\vec{b}$ respectively.

The following should be mentioned in connection with Eqs.(2) and (3):
Many of the well-known phenomena related to Frauenhofer diffraction
have been deduced\cite{Landau} from these equations under the additional
condition (which is directly related to the boundary conditions
imposed on the scattered field)
$|\vec{k}^\prime|=|\vec{k}|=\omega^\prime=\omega$, 
that is,  $\vec{k}^\prime$ differs 
from $\vec{k}$ only in direction.
In other words, the outgoing light wave has exactly the same frequency, 
and exactly the same magnitude of wave-vector
as those for the 
incoming wave. (This means, quantum mechanically speaking, 
the outgoing photons are also on-shell photons, the
energies of which are the same as the incoming ones.)
In such cases, it is possible to envisage that
$\vec{q}$ is approximately perpendicular to $\vec{k}$ and to 
$\vec{k}^\prime$, that is, $\vec{q}$
is approximately
perpendicular to the chosen $z$-axis, and thus in the above-mentioned
$xy$-plane (that is $\vec{q}\approx\vec{q}_\perp$).
While the scattering angle distribution in such 
processes (which are considered as the characteristic features of
{\em elastic}
diffractive scattering) 
plays a significant role in understanding 
the observed diffraction phenomena, it
is of considerable importance to note the following:

(A) Eqs.(2) and (3) can be used 
to describe diffractive scattering
with, or without, this additional
condition, provided that the difference of 
$\vec{k}^\prime$ and $\vec{k}$ in the 
longitudinal direction (i.e. in the direction
of $\vec{k}$) is small compared to 
$q_{\perp}\equiv |\vec{q}_{\perp}|$ 
so that 
$\vec{q}_\perp$ can be approximated by $\vec{q}$. In fact, Eqs.(2) and 
(3) are strictly valid when $\vec{q}$ is a vector in the
above-mentioned $xy$-plane, that is when we write $\vec{q}_\perp$
instead of $\vec{q}$. Now, since Eqs.(2) and (3) in such a form 
(that is when the replacement $\vec{q}\rightarrow \vec{q}_\perp$ is made)
are valid {\em without} the condition $\vec{q}$ should approximately
be equal to $\vec{q}_\perp$ and in particular without the additional condition 
$|\vec{k}^\prime|=|\vec{k}|=\omega^\prime=\omega$,
it is clear that they are also valid for {\em inelastic}
scattering processes.
In other words, Eqs.(2) and (3) can also be used to describe
{\em inelastic}
diffractive scattering (that is, processes
in which 
$\omega^\prime\neq\omega$,
$|\vec{k}^\prime|\neq|\vec{k}|$)
provided that the following replacements are made.
In Eq.(2), $\vec{q}\rightarrow \vec{q}_\perp$,
$f(\vec{q})\rightarrow 
f_{\mbox{\tiny inel.}}(\vec{q}_\perp)$,
$\alpha(\vec{b})\rightarrow
\alpha_{\mbox{\tiny inel.}}(\vec{b})$;
and in Eq.(3)
$q\rightarrow q_\perp$,
$f(q)\rightarrow 
f_{\mbox{\tiny inel.}}(q_\perp)$,
$\alpha(b)\rightarrow
\alpha_{\mbox{\tiny inel.}}(b)$.
Hereafter, we shall call Eqs.(2) and (3) with these replacements
Eq.(2$^\prime$) and Eq.(3$^\prime$)  respectively.
We note, in order to specify the dependence of 
$f_{\mbox{\tiny inel.}}$
on $\omega^\prime$ and $k_{\|}^\prime$
(that is on 
 $\omega^\prime-\omega$ and $k_{\|}^\prime-k_{\|}$),
further information on
energy-momentum transfer 
in such scattering processes is
needed. This point will be discussed in more detail
in Section \ref{sec:3}.

(B) In scattering processes at large momentum-transfer  
where the magnitude of
$\vec{q}_{\perp}$
is large ($|\vec{q}_{\perp}|^2\gg 0.05\,\mbox{GeV}^2$, say), it
is less probable
to find diffractive scattering
events in which the additional condition 
$|\vec{k}^\prime|=|\vec{k}|$ and $\omega^{\prime}=\omega$
can be satisfied. 
This means, it is expected that most of the diffraction-phenomena
observed in such processes are associated with inelastic diffractive 
scattering.

(C) Change in angle but no change in magnitude of wave-vectors or
frequencies is likely to occur in processes in which neither
absorption nor emission
of light takes place. Hence, it is not difficult to imagine, that the 
above-mentioned condition can be readily satisfied in
cases where the scattering systems are time-independent
macroscopic apertures or objects. But, 
in this connection, we are also forced
to the question:
``How large is the chance for a incident wave {\em not} to change the
magnitude of its wave-vector
in processes in which the scatterers are {\em open dynamical
complex systems}, where energy- and momentum-exchange 
take place at anytime and everywhere?!''

The picture for inelastic diffractive scattering
has two basic ingredients:
 
First, having the 
well-known phenomena associated with Frauenhofer's
diffraction
and the properties of 
de Broglie's
matter waves in mind, the beam particles 
($\gamma^\star$, $\gamma$, $\overline{p}$ 
or $p$ shown in Fig.8 of Part I) 
in these scattering processes
are considered as high-frequency 
waves passing through a medium. Since, in general, energy- and
momentum-transfer take place during the passage through the medium,
the wave-vector of the outgoing wave differs, in general, from the
incoming one, not only in direction, but also in magnitude. For the
same reason, the frequency 
and the longitudinal component of the wave-vector
of the outgoing wave (that is the energy, and/or the 
invariant mass, as well as the longitudinal momentum
of the outgoing particles) can be different from their
incoming counterparts.

Second, according to the results obtained in Part I (and summarized in 
the Introduction of Part II) of this paper, the medium is
a system of color-singlet 
gluon-clusters which are in general partly inside and partly outside
the proton --- in form of a ``cluster cloud''\cite{SOCI}. 
Since the average binding energy between such color-singlet aggregates are
of Van der Waals type\cite{Gottfried}, 
and thus it is negligibly small  compared with
the corresponding binding 
energy between colored objects, we expect to see that, 
even at  relatively small values of 
momentum-transfer ($|t|$$<$$1\,\mbox{GeV}^2$,say),
the struck colorless clusters can 
unify with (be absorbed by) the beam-particle, and 
``be carried away'' by the latter,
similar to the process of
``knocking out nucleons'' from nuclear targets in high-energy
hadron-nucleus collisions. It
should, however, be emphasized 
that, in contrast to the nucleons in nucleus, the colorless
gluon-clusters which can exist inside or outside the
confinement-region of the proton
are {\em not} hadron-like (See Sections 3 - 6
of Part I for more details).
They are BTW-avalanches which
have neither a typical size, nor a typical 
lifetime, nor a given static structure. Their 
size- and lifetime-distributions obey simple power-laws
as consequence
of SOC. This means, in the diffraction processes 
discussed here, the size of the scatterer(s), 
and thus the size of the carried-away colorless gluon-cluster(s),
is  in general different in 
every scattering event. 
It should also be emphasized that these
characteristic features
of the scatterer are consequences of
the basic properties of the gluons.

\section{Can such scattering systems be modeled quantitatively?}
\label{sec:3}
To model the proposed picture quantitatively, it is convenient
to consider the scattering system
in the rest frame 
of the proton target. Here, we
choose a right-handed Cartesian coordinate with its origin $O$ at 
the center of the target-proton, 
and the $z$-axis in the direction of the incident 
beam. 
The $xy$-plane in this coordinate system coincides
with the two-dimensional impact-parameter 
space mentioned in connection with Eqs.(2$^\prime$) and (3$^\prime$)
[which are respectively Eq.(2) and Eq.(3) after the replacements
mentioned in (A) below Eq.(3)],
while the $yz$-plane is the 
scattering plane. We note, since 
we are dealing with inelastic scattering (where the momentum 
transfer,  including its component in the longitudinal direction, 
can be large; in accordance with
the uncertainty principle) it is possible 
to envisage that (the c.m.s.\,of) the incident particle in the beam
meets colorless gluon-clusters at one point $B\equiv(0,b,z)$.
where the projection of $\overline{OB}$ along the $y$-axis characterizes
the corresponding impact parameter 
$\vec b$.
We recall that such clusters
are avalanches 
initiated by local perturbations (caused by
local gluon-interactions associated with 
absorption or emission of one or more gluons;
see Part I for details) of SOC states 
in systems of interacting soft gluons. Since gluons 
carry color, the interactions which lead to the formation of
{\em colorless} gluon-clusters  must take place 
inside the confinement region of  the proton. This means, while a 
considerable part of such colorless clusters in the
cloud  can be outside
the proton, the location $A$, where such an avalanche is initiated, 
{\em must} be {\em inside} the proton. That is, in terms of 
$\overline{OA}\equiv r$, $\overline{AB}\equiv R_A(b)$, and  
proton's radius $r_p$,
we have $r\le r_p$ and 
$[R_A(b)]^2=b^2+z^2+r^2-2(b^2+z^2)^{1/2} r \cos\angle BOA$. 
For a given impact parameter
$b$, it is useful to know the distance $R_A(b)$ between $B$ and $A$,
as well as  ``the average squared distance''
$\langle R_A^2(b)\rangle =  b^2+z^2 + a^2$,
$a^2\equiv 3/5\,r_p^2$,
which is
obtained by averaging  over all allowed locations of $A$
in the confinement region.
That is, we can model {\em the effect of confinement} in 
cluster-formation by picturing that all the avalanches,
in particular those
which contribute to 
scattering events characterized by a given $b$ and a given $z$ 
are initiated from an ``effective initial point'' $\langle
A_B\rangle$,
because only the mean distance between $A$ and $B$ 
plays a role.
(We note, since we are dealing with a complex system with many degrees 
of freedom, in which
$B$ as well as $A$ are randomly chosen points in
space, 
we can compare {\em the mean distance} between $B$ and $A$ 
with the mean free path in a gas mixture
of two kinds of gas molecules ---
``Species $B$'' and ``Species $A$'' say, where those of the latter kind are
confined inside a subspace called ``region $p$''.
For {\em a given mean distance}, and {\em a given point $B$},
there is in general a set of $A$'s inside the ``region $p$'',
such that their distance to $B$ is {\em equal} to the given mean
value. Hence it is useful to introduce a {\em representative} point
$\langle A_B\rangle$, such that the distance
between $\langle A_B\rangle$ and $B$ is equal 
to the given mean distance.)
Furthermore, since an avalanche is a dynamical object, it may propagate 
within its lifetime
in any one of the $4\pi$
directions away from $\langle A_B\rangle$. (Note: avalanches of
the same size may have different lifetimes, different structures, 
as well as 
different shapes. The location of an avalanche
in space-time is referred to its center-of-mass.)
Having seen how SOC and confinement can be implemented in describing
the properties and the dynamics of colorless gluon-clusters, 
which are nothing else but BTW-avalanches in systems of interacting
soft gluons,
let us now go one step further, and discuss how these
results can be used to obtain the amplitudes in impact-parameter-space 
that leads, via
Eq.(3), to the scattering amplitudes.

In contrast to the usual cases, where the scatterer  
in the optical
geometrical picture of a diffractive scattering process
is an aperture, or an 
object, with a given static structure,
the scatterer in the proposed picture is
an open dynamical complex system of
 colorless gluon-clusters in form of BTW-avalanches.
This implies in particular: The object(s), which the beam particle 
hits, has (have) neither a typical size, nor a typical lifetime, nor a
given static structure.

With these in mind, let us now come back to our discussion on the
double differential cross section $d^2\sigma/dt\,d(M_x^2/s)$.
Here, we need to determine
the corresponding amplitude 
$\alpha_{\mbox{\tiny inel.}}(b)$
in Eq.(3$^\prime$) [see the discussion in (A) below Eq.(3)]. 
What we wish to do now, is to 
focus our attention on those
scattered matter-waves whose de Broglie wavelengths are
determined by the energy-momentum of the scattered 
object, whose invariant mass  is $M_x$.
For this purpose, we characterize
the corresponding $\alpha_{\mbox{\tiny inel.}}(b)$ 
by considering it as a function of
$M_x$, or $M_x^2/s$, or $x_P$. We 
recall in this connection that, for inelastic diffractive 
scattering processes in hadron-hadron collisions, the quantity $M_x^2/s$
is approximately equal to $x_P$, which is 
the momentum fraction carried by the struck
colorless gluon-cluster with respect  to the incident beam
(see Fig.8 of Part I for more details; note however
that $q_c$, $k$ and $p_x$ in Fig.8 of
part I correspond respectively to
$q$, $k$ and $k^\prime$ in the discussions here.).
Hence, we shall write hereafter 
$\alpha(b\,|M_x^2/s)$ or $\alpha(b\,|x_P)$
instead of the general expression
$\alpha_{\mbox{\tiny inel.}}(b)$.
This,
together with Eq.(3$^\prime$), leads to the corresponding
scattering amplitude
$f(q_\bot |x_P)$, and 
thus to the corresponding double differential cross-section
$d^2\sigma/dt\,dx_P$,
in terms of the variables
$|t|\approx |\vec{q}_{\bot}|^2$ 
and $x_P\approx M_x^2/s$ in the kinematical region: $|t|\ll M_x^2 \ll s$.

\section{The role played by the space-time properties of the
gluon-clusters}
\label{sec:4}
For the determination of $\alpha(b\,|x_P)$, it is of considerable
importance to recall the following space-time properties of the
color-singlet gluon-clusters which are BTW-avalanches due to SOC:

(i) SOC dictates, that there are BTW-avalanches of all sizes 
(which we denote by different $S$ values), and
that the probability amplitude of finding an avalanche of size
$S$ can be obtained from the size-distribution
$D_S(S)=S^{-\mu}$ where the  
experimental results
presented in Part I
 show: $\mu\approx 2$.
This means, $D_S(S)$ contributes a factor 
$S^{-1}$, thus a factor 
$x_P^{-1}$ to the scattering amplitude $\alpha(b\,|x_P)$.
Here, as well as in (ii), we take into account
(see Sections 4 and 5 of Part I for details) 
that the size
$S$ of a colorless gluon-cluster is directly proportional to the
total amount of the energy the cluster carries;
the amount of energy is $x_P P^0$,
where $P^0$ is the total energy of the proton, and $x_P$ is the 
energy fraction carried by the cluster.

(ii) QCD implies\cite{Gottfried} that the 
interactions between two arbitrarily
chosen colored constituents of 
a BTW-avalanche (which is a colorless gluon cluster) is stronger than those 
between two color-singlet BTW-avalanches, 
because the latter should be interactions of Van der
Waals type. This means, the struck avalanche can 
unify with the beam-particle (maybe by absorbing each other), and
viewed from any Lorentz frame in which the beam-particle has a larger
momentum than that of the colorless gluon-cluster, the latter is
``carried away'' by the beam particle.
Geometrically, the chance for the 
beam-particle to hit an avalanche
of size $S$ ( on the plane perpendicular to the incident axis) 
is proportional to  the area that can be struck by the (c.m.s.) of the 
beam particle. The area is the $2/3$-power of the volume $S$,
$S^{2/3}$,
and thus it is proportional to $x_P^{2/3}$.

(iii) Based on the above-mentioned picture 
in which the BTW-avalanches propagate isotropically
from $\langle A_B\rangle$,  the
relative number-densities at different $b$-values can be readily 
evaluated. Since for a given $b$, the distance
in space between $\langle A_B\rangle$ and
$B\equiv (0,b,z)$ is simply 
$(b^2+z^2+a^2)^{1/2}$, the number of avalanches which pass a unit area 
on the shell of radius $(b^2+z^2+a^2)^{1/2}$ centered at $\langle A_B\rangle$
is proportional to $(b^2+z^2+a^2)^{-1}$, provided that (because of causality)
the lifetimes ($T$'s) of these avalanches are 
not shorter than
$\tau_{\mbox{\tiny min}}(b)$. The latter is the time interval for an
 avalanche to travel from  $\langle A_B\rangle$ to $B$.
This means, because of the space-time properties of such avalanches it 
is of considerable importance to note: 
First, only  avalanches having lifetimes
$T\ge\tau_{\mbox{\tiny min}}(b)$ can contribute to such a collision
event. Second, during the propagation from $\langle A_B\rangle$ to
$B$, the motion of such an avalanche has to be considered as
Brownian. In fact, the continual, and more or less random, impacts
received from the neighboring objects on its path leads us to the well 
known \cite{StatMech} result that the time elapsed is proportional to
the mean-square displacement. 
That is: $\tau_{\mbox{\tiny min}}(b)\propto b^2+z^2+a^2$.
Furthermore, we recall that avalanches are due to SOC, 
and thus the chance for an avalanche
of lifetime  
$T$ to exist is 
$D_T(T) \propto T^{-\nu}$ 
where the experimental value\cite{SOCI} for
$\nu$ is $\nu\approx 2$. Hence, by integrating 
$ T^{-2}$ over $T$ from $\tau_{\mbox{\tiny min}}(b)$ to infinity,
we obtain
the fraction associated with all those whose lifetimes satisfy $T\ge
\tau_{\mbox{\tiny min}}(b)$: This fraction 
is $\tau_{\mbox{\tiny min}}(b)^{-1}$, and thus
a constant times  $(b^2+z^2+a^2)^{-1}$.

The amplitude $\alpha(b\,|x_P)$ can now be obtained from
the probability amplitude for avalanche-creation
mentioned in (i), by 
taking the weighting factors mentioned in
(ii) and (iii) into account, and by integrating over $z$ \cite{zInt}. 
The result is:
\begin{eqnarray}
\alpha(b\,|x_P) = \mbox{const.} x_P^{-1/3}(b^2+a^2)^{-3/2}\mbox{\,.}
\end{eqnarray}
By inserting this
probability amplitude in
impact-parameter-space, for the beam particle to encounter colorless
gluon-clusters (avalanches in the BTW-theory) which carries a
fraction $x_P$ of the proton's total energy, in  
Eq.(3') [which is Eq.(3) with the folowing replacements:
$q\rightarrow q_\perp$,
$f(q)\rightarrow f_{\mbox{\tiny inel.}}(q|x_P)$ and
$\alpha(b)\rightarrow \alpha_{\mbox{\tiny inel.}}(b|x_P)$]
we obtain
the
corresponding probability amplitude $f(q|x_P)$ in 
momentum-space:
\begin{equation}
f(q_\bot |x_P)=\mbox{const.}\int_0^{\infty} 
b\, db\,x_P^{-1/3}(b^2+a^2)^{-3/2}J_0(q_\bot b),
\end{equation}
where $q_\bot =|\vec{q}_\bot |\approx \sqrt{|t|}$ 
(in the small $x_P$-region, $x_P<0.1$, say)
is the 
corresponding momentum-transfer.
The integration can be carried out analytically\cite{Integrals},
and the result is 
\begin{equation}
f(q_\bot |x_P)={\rm const.}x_P^{-1/3}\exp (-a q_\bot )\mbox{\,.}
\end{equation}
Hence, the corresponding double differential cross-section
$d^2\sigma/ dt dx_P$ can approximately be written as
\begin{equation}
\frac{1}{\pi}{d^2\sigma\over dt\,dx_P}=N x_P^{-2/3}\exp (-2a\sqrt{|t|}),
\end{equation}
where $N$ is an unknown normalization constant. Because of the
kinematical relationship $x_P\approx M_x^2/s$ for 
single diffractive scattering in proton-proton and
proton-antiproton collisions
(see Fig.8 of Part I for more details),
this can be, and
should be, compared with the measured double differential
cross-sections $d^2\sigma/dt\,d(M_x^2/s)$ at different $t$- and
$s$-values and for different missing masses $M_x$
in the region $M_x^2/s\ll 1$ 
where $q_\perp$ is approximately $\sqrt{|t|}$. 
The comparison is shown
in Fig.1. Here, we made use of the fact that $a^2\equiv 3/5\,r_p^2$, 
where $r_p$ is the proton radius, and calculated $a$ by setting $r_p^2$
to be the 
well-known \cite{Halzen} 
mean square proton charge radius, the value of which is
$r_p^2 = (0.81\,\mbox{fm})^2$.
The result we obtained is:
$a= 3.2\,\mbox{GeV}^{-1}$. The unknown normalization constant
is determined by inserting this calculated value for $a$ in Eq.(7), and 
by comparing the right-hand-side of this equation with
the $d^2\sigma/dt\,d(M_x^2/s)$ data taken
at $|t|= 0.2\,\mbox{GeV}^2$. The value is $N=31.1\,\mbox{mb
GeV}^{-2}$. All the curves shown in Fig.1 are obtained by inserting these 
values for $a$ and $N$ in Eq.(7).

While the quality of the obtained result, namely the expression given
on the right-hand-side of Eq.(7)
together with the above-mentioned values for $a$ and $N$,
can be readily judged by comparing
it with the data, or by counting the unknown parameters, or both, it
seems worthwhile to recall the following:
The two basic ingredients of the proposed picture which have been used 
to derive this simple analytical expression are: first, the well-known 
optical analogy, and second, the properties of the dynamical
scattering system. The latter is what we have learned through the
data-analysis presented in Part I of this paper.

Based on the theoretical arguments and experimental indications 
for the observation (see Ref.[\ref{SOCLetter}] and Part I of this paper
for details) that the characteristic features of
inelastic diffractive scattering processes are approximately
independent of the incident energy and independent of the
quantum-numbers of the beam-particles, the following
results are expected:
The explicit formula for the double differential cross-section 
as shown in Eq.(7) should also be valid for the reactions 
$\gamma p\rightarrow X p$ and 
$\gamma^\star p\rightarrow X p$.
While the normalization constant $N$ (which should in particular
depend on the geometry of the beam particle) is expected to be
different for different reactions, everything else --
especially the ``slope'' as well as the power of $x_P$ should be
exactly the same as in $pp$- and $p\bar{p}$-collisions.
In this sense, Eq.(7) with $a^2=3/5r_P^2$ ($r_P$ is the proton radius) 
is our prediction for $\gamma p \rightarrow X p$ and $\gamma^\star p
\rightarrow X p$ which can be measured at HERA.

Furthermore, in order to obtain the integrated differential cross-section
$d\sigma/dt$, which has also been measured for different
reactions at different incident energies, we only need to
sum/integrate over $x_P$ in the given kinematic range ($x_P<0.1$,
say). The result is
\begin{equation}
{d\sigma\over dt}(t)=C\exp (-2a\sqrt{|t|}),
\end{equation}
where $C$ is an unknown normalization
constant. 
While this observation
has already been briefly discussed in a
previous note \cite{SOCLetter}, 
we now show the result of a further 
test of its universality: In Fig. 2, we plot
\begin{equation}
-{1\over 2\sqrt{|t|}}\log [\frac{1}{C}\,{d\sigma\over dt}(t)] \mbox{ vs. } t 
\end{equation}
{\em for different reactions at different incident energies} 
in the range 
$0.2$$\mbox{\,GeV}^2$$\le$$|t|$$\le 4\mbox{\,GeV}^2$.
Here we see in particular that, 
measurements of $d\sigma/dt$ for $\gamma^{\star}p$ and
$\gamma p$ reactions at larger $|t|$-values would be very useful.

\section{Concluding remarks}
\label{sec:5}
Based on the characteristic properties of the gluons ---
in particular the local gluon-gluon coupling prescribed by
the QCD-Lagrangian, the confinement, and the non-conservation 
of gluon-numbers, we suggest that a system of interacting
soft gluons should be considered as an open dynamical complex system
which is in general far away from equilibrium.
Taken together with the observations made 
by Bak, Tang and Wiesenfeld (BTW) \cite{OriginalSOC,ReviewSOC},
we are led to the conclusion\cite{SOCI}, 
that self-organized criticality (SOC) and 
thus BTW-avalanches exist in such systems, and that 
such avalanches 
manifest themselves in form of color-singlet gluon-clusters 
in inelastic diffractive scattering processes.

In order to test this proposal, we performed a systematic
data-analysis, the result of which is presented in Part I of this
paper: It is shown that the size-distributions, 
and the lifetime-distributions, of
such gluon-clusters {\em indeed} exhibit power-law behaviors which are known 
as the fingerprints of SOC \cite{OriginalSOC,ReviewSOC}. 
Furthermore, it is found that such exponents
are approximately the same for different reactions at different
incident energies --- indicating the expected universality and
robustness of SOC. Hence, the following picture emerges:
For the beam particle (which may be a 
virtual photon, or a real photon, or 
a proton, or an antiproton; see Fig.8 in Part I for more details) 
in an inelastic diffractive scattering process 
off proton (one may wish to view this from a ``fast moving frame''
such as the c.m.s. frame), 
the target proton appears as a cloud of colorless
gluon-clusters which exist inside and outside the confinement
region of the proton. The size (S) distribution $D_S(S)$ and the
lifetime
distribution $D_T(T)$ can be expressed as $S^{-\mu}$ and
$T^{-\nu}$ respectively, where 
the empirical values for $\mu$ and $\nu$ are
$\mu$$\approx$$\nu$$\approx$$2$,
independent of the incident energy, and independent of the quantum
numbers of the beam particles.

What do we learn from this? Is this knowledge
helpful in understanding hadronic structure and/or
hadronic reactions 
in Particle Physics? In particular, can 
this knowledge be used to do {\em quantitative} 
calculations --- especially those, the results of
which could {\em not} be
achieved otherwise?

In order to demonstrate how the obtained knowledge can be used to 
relate hadron-structure and hadronic reactions in general, and to
perform quantitative calculations in particular, we discuss the
following question --- a question which has been with the high-energy 
physics community for many years:

``Can the measured double differential cross section 
$d^2\sigma/dt\,d(M_x^2/s)$ 
for 
inelastic diffractive scattering in
proton-proton and in antiproton-proton collisions,
in the kinematical region given by 
$0.2\mbox{\,GeV}^2$$\le$$|t|$$\le$ $3.25 \mbox{\,GeV}^2$, 
$16\mbox{\,GeV}$$\le$$\sqrt{s}$$\le$ $630\mbox{\,GeV}$,
and $M_x^2/s$ $\le$$0.1$, be understood in terms of optical
geometrical concepts?'' 

The answer to this question is ``Yes!'', and the 
details are presented in Sections \ref{sec:2}, \ref{sec:3} and
\ref{sec:4} of Part II where
the following are explicitly shown:
The characteristic features of the existing
$d^2\sigma/dt\,d(M_x^2/s)$-data are very much the same as those in
optical diffraction, provided that the high-energy beams are
considered as high-frequency waves, and the scatterer is a
system of colorless gluon-clusters described in Part I of this paper.
Further measurements of double differential cross sections, especially
in $\gamma^{\star}p$- and $\gamma p$-reactions, will be helpful in
testing the ideas presented here.

\subsection*{Acknowledgement}
We thank C.B. Yang and W. Zhu for helpful discussions, and FNK der FU
Berlin for financial
support. Y. Zhang thanks Alexander von Humboldt Stiftung for the
fellowship granted to him.

\newpage

\begin{figure}[hbt]
\label{figure1}
\caption{The double 
differential cross section $(1/\pi)\,d^2\sigma/ dt\,d(M_x^2/s)$ 
for single diffractive $pp$ and $\bar{p}p$ reactions
is shown as function of
 $x_P$ at fixed values of $t$ where
$0.15$$\mbox{\,GeV}^2$$\le$$|t|$$\le$$3.25$$\mbox{\,GeV}^2$.
The data are taken from Refs.
[\ref{ppdata}-\ref{hhdata}]. The solid curve is the
result obtained from Eq.(7). 
The dashed curve stands for the result obtained from the same formula
by using the $t$-value given in the bracket.}
\end{figure}

\begin{figure}[hbt]
\label{figure2}
\caption{
The quantity $(-1/(2 \sqrt{|t|})) \log{[\frac{1}{C}\,d\sigma/ dt]}$ 
is plotted versus $\sqrt{|t|}$ for different single diffractive
reactions
in the range
$0.2$$\mbox{\,GeV}^2$$\le$$|t|$$\le$$4$$\mbox{\,GeV}^2$.
The data are taken from Refs. [\ref{LRGDiscovery}-\ref{hhdata}].
Here, $C$, the normalization constant is first
determined by performing a two-parameter fit of 
the corresponding $d\sigma/dt$-data to Eq.(8).}
\end{figure}

\newpage

\psfig{figure=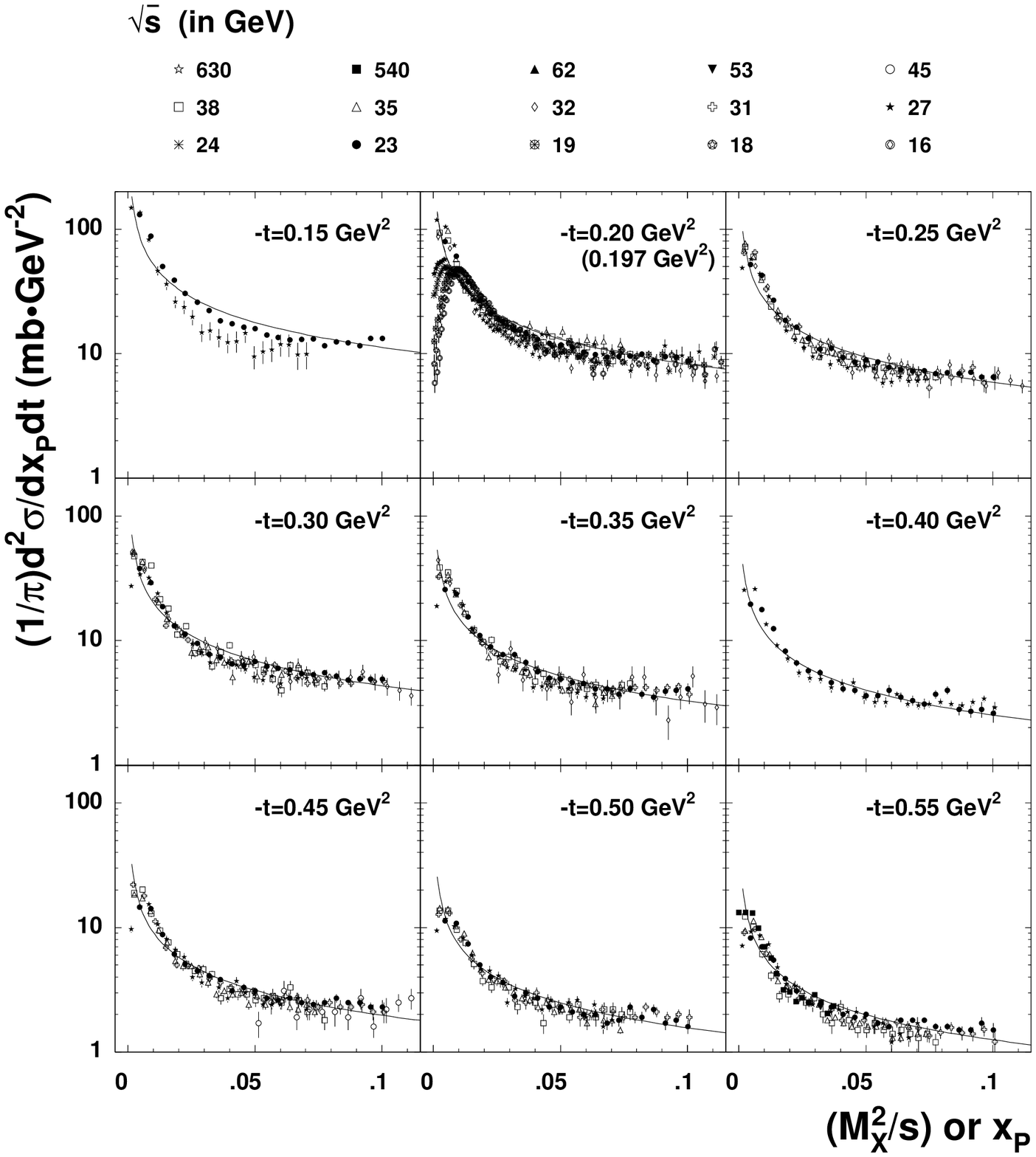,width=16.5cm}

\psfig{figure=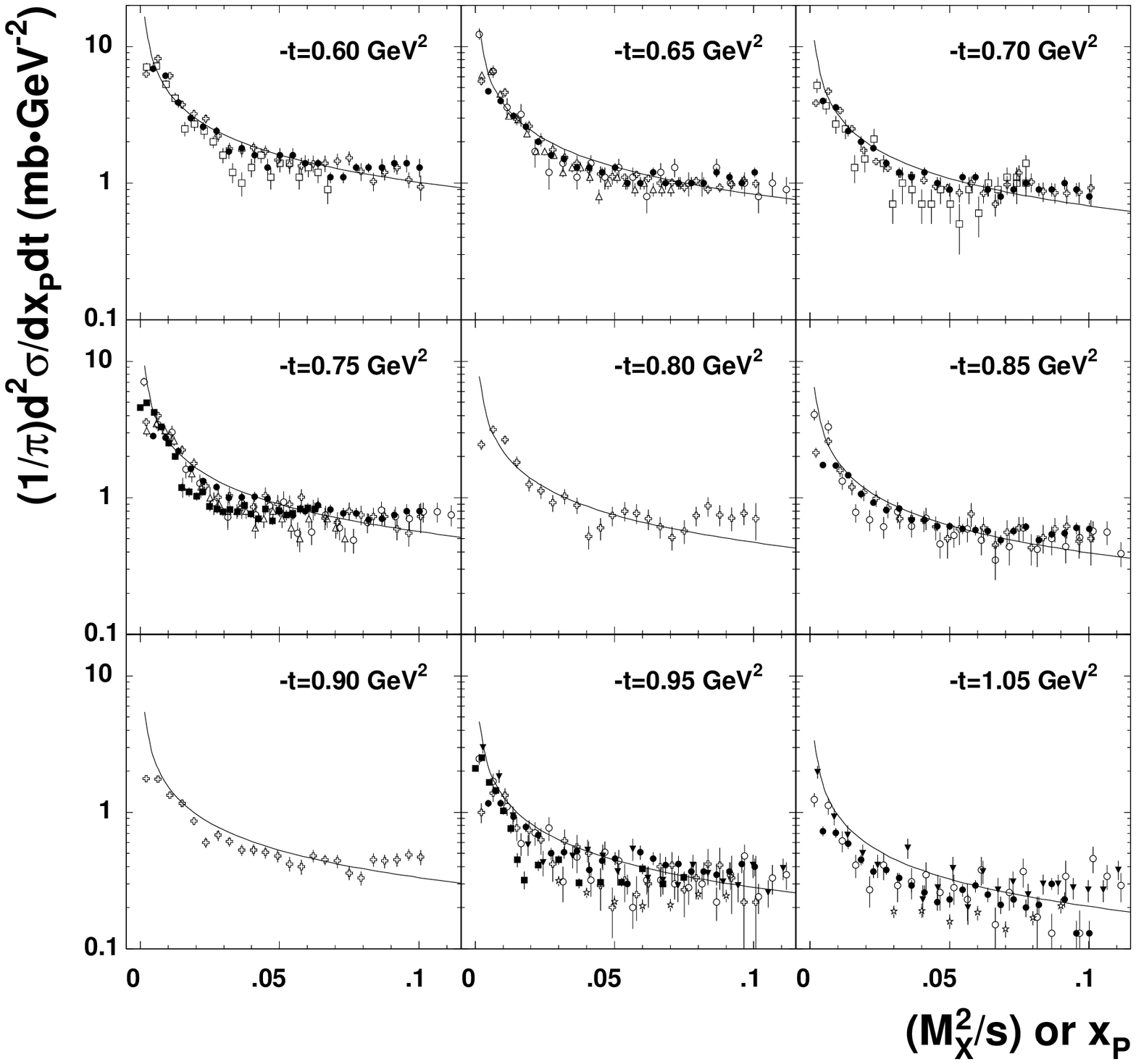,width=16.5cm}

\psfig{figure=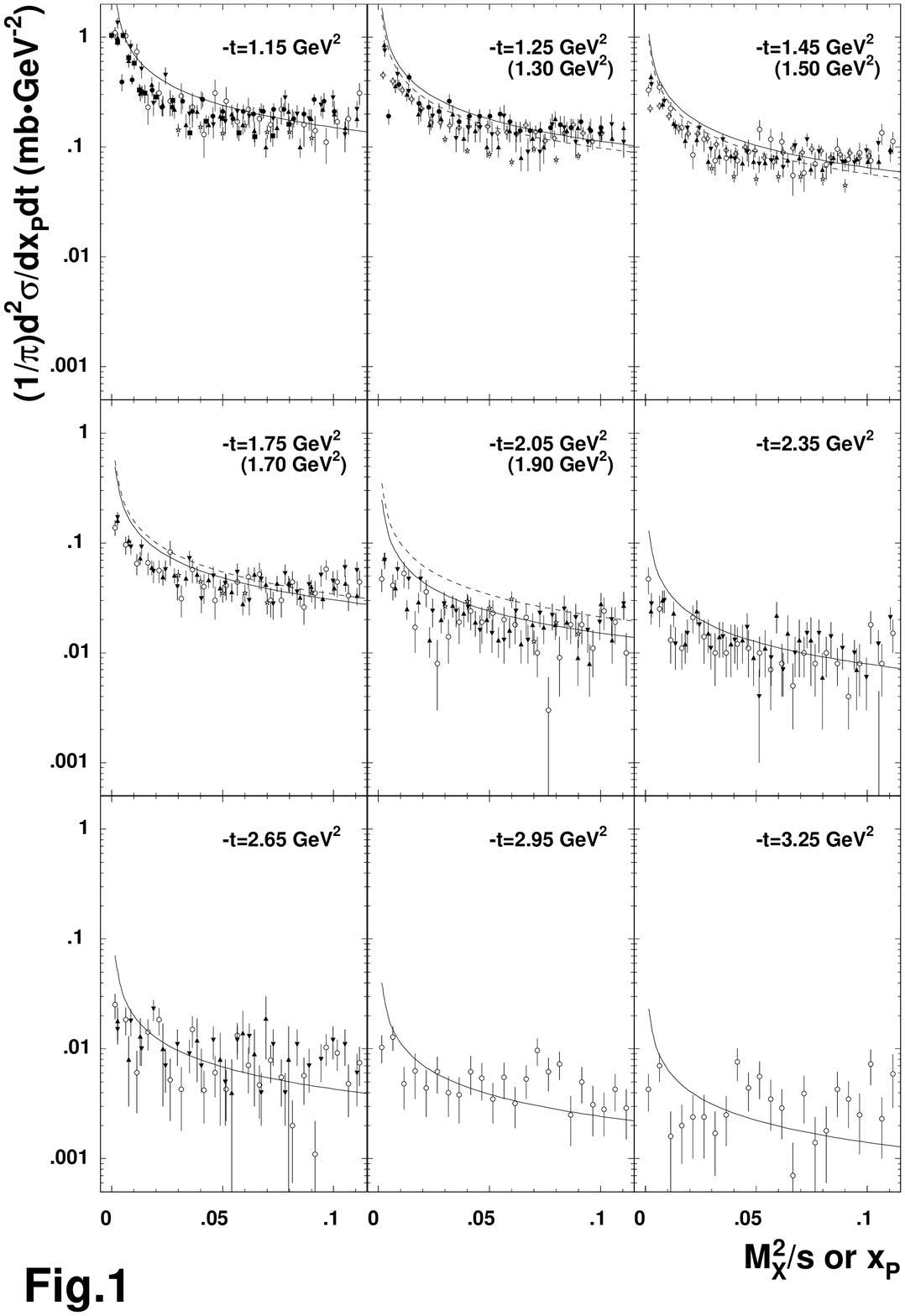,width=16.5cm}

\psfig{figure=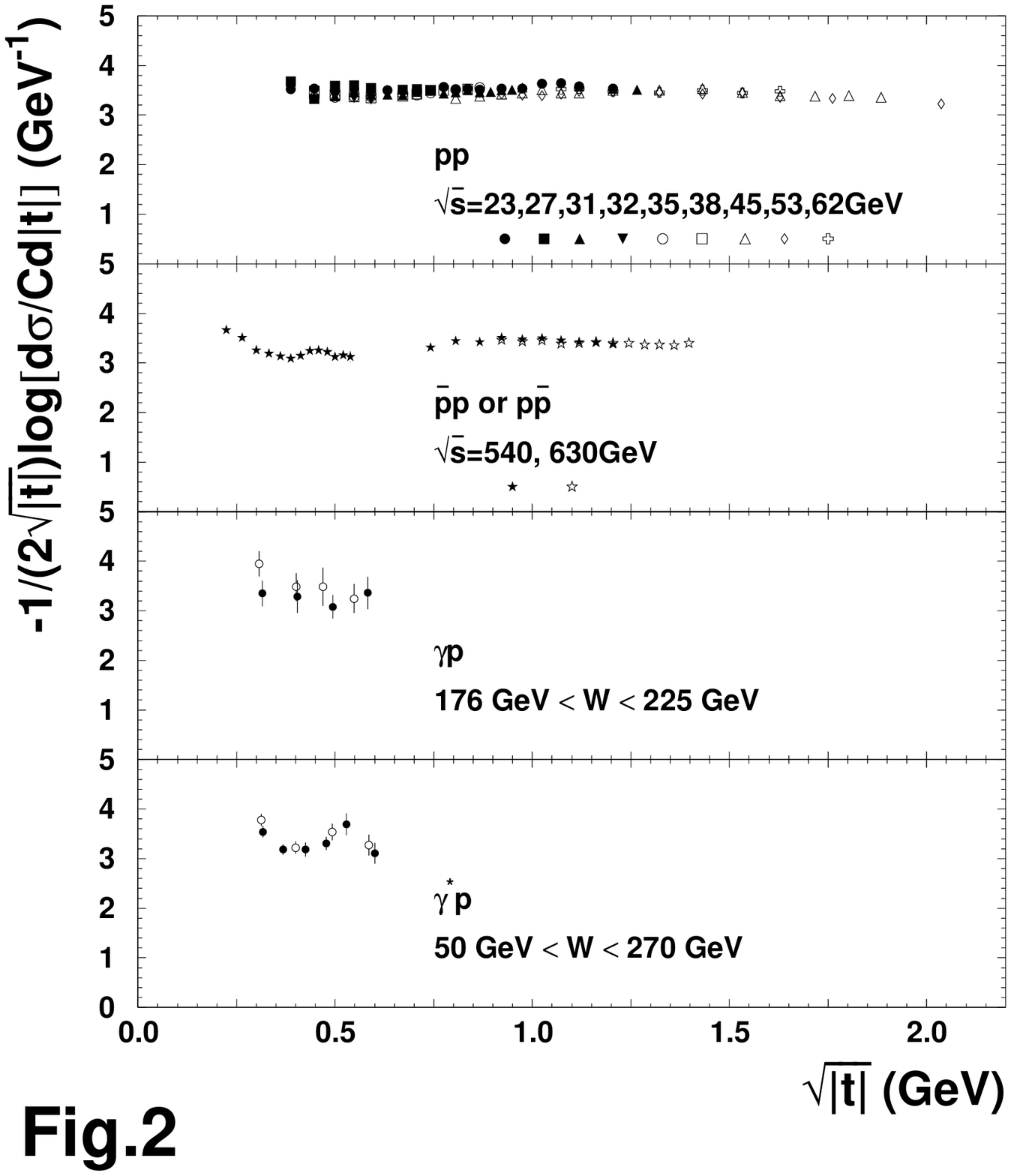,width=16.5cm}


\begin{thebibliography}{99}
\bibitem{SOCI}
\label{SOCI}
C. Boros, T. Meng, R. Rittel and Y. Zhang, 
Part I of this paper.

\bibitem{LRGDiscovery} 
\label{LRGDiscovery}
M. Derrick, et al., ZEUS Coll.,  Phys. Lett. {\bf B315}, 481 (1993);
T. Ahmed, et al., H1 Coll.,  Nucl. Phys. {\bf B429}, 477 (1994).

\bibitem{LRGDISatHera} 
\label{LRGDISatHera}
M. Derrick, et al., ZEUS Coll., Phys. Lett. {\bf B345}, 576 (1995);
M. Derrick, et al., ZEUS Coll.,  Z. Phys. {\bf C68}, 569 (1995);
{\rm ibid}, {\bf C70}, 391 (1996);
T. Ahmed, et al., H1 Coll.,  Phys. Lett. {\bf B348}, 681 (1995);
J.Breitweg, et al., ZEUS Coll., Eur. Phys. J. {\bf C1}, 81 (1998).


\bibitem{gammapdata}
\label{gammapdata}
C. Adloff, et al., H1 Coll., Z. Phys. {\bf C74}, 221 (1997);
J. Breitweg, et al., ZEUS Coll., Z. Phys. {\bf C75}, 421 (1997);
M. Derrick, et al., ZEUS Coll., Phys. Lett. {\bf B293}, 465 (1992).

\bibitem{ppdata}
\label{ppdata}
M.G. Albrow, et al., CHLM Coll.,  Nucl. Phys. {\bf B108}, 1 (1976); 
R. D. Schamberger, et al., Phys. Rev. {\bf D17}, 1268 (1978).

\bibitem{ppbardata}
\label{ppbardata}
M. Bozzo, et al., UA4 Coll., Phys. Lett. {\bf B136}, 217 (1984);
D. Bernard, et al., UA4 Coll.,   Phys. Lett. {\bf B186}, 227 (1987);
A. Brandt, et al., UA8 Coll.,  Nucl. Phys. {\bf B514}, 3 (1998).


\bibitem{hhdata} 
\label{hhdata}
R. L. Cool, et al., Phys. Rev. Lett. {\bf 47}, 701 (1981).

\bibitem{GluonDominance}
\label{GluonDominance}
See for example: 
M. Derrick, et al., ZEUS Coll., Phys. Lett. {\bf B345}, 576 (1995);
S. Aid, et al., H1 Coll., Phys. Lett. {\bf B354}, (1995) 494; 
and the references cited therein.

\bibitem{OriginalSOC} 
\label{OriginalSOC}
P. Bak, C. Tang and K. Wiesenfeld, Phys. Rev. Lett. {\bf 59}, 381 (1987); 
Phys. Rev. {\bf A38}, 364 (1988).

\bibitem{ReviewSOC}
\label{ReviewSOC}
See for example:
P. Bak and M. Creutz, in
{\it Fractals and Self-organized Criticality},  
in {\it Fractals in Science}, edited by A. Bunde and S. Havlin
(Springer, Berlin, Heidelberg 1994);
P. Bak, {\it How nature works} 
(Springer, New York 1996); and the references therein.

\bibitem{SmalltNew}
\label{SmalltNew}
See  for example: J. Bartels, in {\it Proceedings of the
 17th Int. Symp. on Lepton-Photon Interactions,
Beijing, 1995}, edited by Zheng Zhi-peng and Chen He-sheng, 
(World Scientific, 1996), Vol. 2, p. 554;
H. Abramowicz, J. Bartels, L. Frankfurt and H. Jung,
in  {\it Proceedings of the Workshop on Future Physics at
HERA}, edited by G. Ingelman, A. De Roeck, R. Klanner, (DESY, 1996), 
Vol. 2, p. 535; 
E. Gallo, in {\it Proceedings of the 18th Int.
Symp. on Lepton - Photon Interactions, Hamburg, 1997}, edited by
A. DeRoeck and A. Wagner, (World Scientific, Singapore 1998); 
and the references therein.


\bibitem{softgluon}
\label{softgluon}
F. E. Low, Phys. Rev. {\bf D12}, 163 (1975);
S. Nussinov, Phys. Rev. Lett. {\bf 34}, 1286 (1975);
C. Boros, Z. Liang and T. Meng, Phys. Rev. {\bf D54}, 6658 (1996).

\bibitem{ReviewSmallt}
\label{ReviewSmallt}
See for example:
M. M. Block and R. N. Cahn, Rev. Mod. Phys., Vol. {\bf 57}, No. 2, 563 (1985);
J. Lee-Franzini, in {\it AIP Conference Proceedings No. 15}, (American
Institute of Physics, New York 1973) p.147;
K. Goulianos, Phys. Rep. {\bf 101} No.3, 169 (1983);
and in {\it Proceedings of DIS97, Chicago, 1997}, edited by
J. Repond and D. Krakauer, (AIP), p. 527;
U. Amaldi, M. Jacob, and G. Matthiae, 
Ann. Rev. Nucl. Sci. {\bf 26}, 385 (1976) and the references 
therein.



\bibitem{OpticalModels}
\label{OpticalModels}
See for example: 
R. Serber, Rev. Mod. Phys. {\bf 36}, 649 (1964);
N. Byers and C. N. Yang, Phys. Rev. {\bf 142}, 976 (1966);
T. T. Chou and C. N. Yang, Phys. Rev. {\bf 170}, 1591 (1968);
{\bf 175}, 1832 (1968); 
{\bf 22}, 610 (1980);
and the references therein.


\bibitem{ReggeModels}
\label{ReggeModels}
See in particular:
F. E. Low, Phys. Rev. {\bf D12}, 163 (1975);
S. Nussinov, Phys. Rev. Lett.{\bf 34}, 1286 (1975);
G. Ingelman and  P. Schlein, Phys. Lett. {\bf B152}, 256 (1985);
A. Donnachie and P.V. Landshoff, Phys. Lett. {\bf B191}, 309 (1987);
Y. A. Simonov, Phys. Lett. {\bf B249}, 514 (1990);
G. Ingelman and  K. Janson-Prytz, Phys. Lett. {\bf B281}, 325 (1992);
G. Ingelman and  K. Prytz, Z. Phys. {\bf C58}, 285 (1993);
and the references therein.

\bibitem{Sterman}
\label{Sterman}
G. Sterman (private communication).

\bibitem{SOCLetter}
\label{SOCLetter}
T. Meng, R. Rittel and Y. Zhang, Phys. Rev. Lett. (in press).

\bibitem{Landau}
\label{Landau}
See for example:
L. D. Landau and E. M. Lifshitz, {\it The Classical Theory of Fields},
2nd rev. edn. (Pergamon Press, Oxford 1962), p. 177 and p. 165.

\bibitem{Gottfried}
\label{Gottfried}
See for example:
K. Gottfried and V. Weiskopf, {\it Concepts in Particle Physics},
(Oxford University Press, New York, and Clarendon Press, Oxford
1986), vol. II, p. 347.

\bibitem{StatMech}
\label{StatMech}
See for example:
 R. K. Pathria, {\it Statistical Mechanics}, (Pergamon Press, Oxford 
1972), p. 451. 

\bibitem{zInt}
\label{zInt}
To be more precise, by 
taking all the mentioned factors into account we obtain
$\alpha(b,z|x_P)\propto \mbox{const }
x_P^{-1+2/3} (b^2+z^2+a^2)^{-1-1}$. Note also that $\int_{-\infty}^\infty 
(b^2+z^2+a^2)^{-2} dz \propto (b^2+a^2)^{-3/2}$.


\bibitem{Integrals}
\label{Integrals}
See for example: 
I. S. Gradshteyn and I. M. Ryzhik, {\it Table of Integrals, 
Series and Products},
(Academic Press, New York 1980), p. 682.

\bibitem{Halzen}
\label{Halzen}
See for example: 
F. Halzen and A. D. Martin, {\it Quarks and Leptons: An
Introductory Course in Modern Particle Physics}, (John Wiley and Sons,
1984), p. 179.


\end{thebibliography}
\end{document}